\documentclass[aps,pre,floatfix,superscriptaddress,twocolumn]{revtex4}
\usepackage{dcolumn}
\usepackage{graphicx}
\usepackage{epsfig}
\usepackage[utf8]{inputenc}
\usepackage[english]{babel}
\usepackage{color}
\newcommand{\gd}{\color{black}}

\newcommand\T{\rule{0pt}{2.6ex}}

\begin{document}

\title{ How curvature flows: scaling laws and global geometry of impact induced attrition processes}

\author{Gerg\H o P\'al}
\affiliation{Department of Theoretical Physics, Doctoral School of Physics, 
Faculty of Science and Technology, University of Debrecen, P.O.\ Box 400, H-4002 Debrecen, Hungary}
\affiliation{Institute of Nuclear Research (Atomki), P.O.\ Box 51, H-4001 Debrecen, Hungary}
\author{G\'abor Domokos}
\affiliation{Department of Mechanics, Materials and Structures, Budapest University of Technology and Economics, 
M\H uegyetem rkp. 3., K261, 1111 Budapest, Hungary}
\affiliation{MTA-BME Morphodynamics Reserarch Group, M\H uegyetem rkp. 3., K261, 1111 Budapest, Hungary}
\author{Ferenc Kun}
\email{Corresponding author: ferenc.kun@science.unideb.hu}
\affiliation{Department of Theoretical Physics, Doctoral School of Physics, 
Faculty of Science and Technology, University of Debrecen, P.O.\ Box 400, H-4002 Debrecen, Hungary}

\date{\today}


\begin{abstract}
Impact induced attrition processes are, beyond being essential models of industrial ore processing, broadly regarded as the key to decipher the provenance of sedimentary particles. 
A detailed understanding of single impact phenomena of solid bodies has been obtained in physics 
and engineering, however, the description of gradual mass reduction and shape evolution  
in impact sequences relies on approximate mathematical models of mean field type, 
formulated as curvature-driven partial differential equations. 
Here we establish the first link between microscopic, particle-based material models and the mean field theory for these processes.
Based on realistic computer simulations of particle-wall collision sequences, we first identify the well-known damage and fragmentation energy phases, then we show that the former is split into the \emph {abrasion phase} with infinite sample lifetime, analogous to Sternberg's Law, at finite asymptotic mass and the \emph{cleavage phase} with finite sample lifetime, decreasing as a power law of the impact velocity, analogous to Basquin's Law. We demonstrate that only in the abrasion phase does shape evolution emerging in microscopic material
models reproduce with startling accuracy the spatio-temporal patterns 
predicted by macroscopic mean field approaches. 
Our results substantially extend the phase diagram of impact phenomena and set the 
boundaries of the applicability of geometric mean field theories for geological shape evolution. 
Additionally, the scaling laws obtained can be exploited for quantitative predictions of evolution histories.
\end{abstract}

\maketitle

\section{Introduction}

Impact induced damage and fragmentation of solids is ubiquitous in nature  and plays
a crucial role in the evolution of our geological environment: repeated impacts
shape particles (sand grains, pebbles, and volcanic rocks) in sediment transport 
\cite{wald_form_1990,lorang_pebble_1990,turcotte_fractals_1997,durian_what_2006,Williams31052013,
doan_rock_2009,dufek_granular_2012,szabo_universal_2018,Domokos2014}, 
affect the production of ash and pyroclast particles in volcanic 
eruptions \cite{jones_ash_2017,hornby_phase_2019}, and contribute to the generation 
of atmospheric aerosols with consequences on air pollution 
and on the global climate \cite{kok_scaling_2011}. In the Solar system, the size and shape 
of asteroids and of the particles of planetary rings observed today are the results 
of a long lasting collisional evolution
\cite{dahmen_nature_2011, farinella_asteroid_1997,Durda201577_fragment_shape, Domokos_2009, Domokos_2017}.
On planet Mars traces of fluvial evolution of landforms such as pebbles have been discovered \cite{Williams31052013,szabo_reconstructing_2015} similar to river beds on Earth.
Particle breakage is widely used by the industry in comminution processes
of ores and minerals 
\cite{salman_classify_impact_2004,WILLS2016109,Spahn_2014,brilliantov_size_2015,andrews_frag_1999}, 
however, it can also be undesired in process and handling
engineering due to the resulting degradation of product quality. 
In these natural processes and industrial applications particles collide both with
each other and with hard walls presented by Earth surface (river beds, beaches, and 
rock walls) or by the components of the process equipment (conveyors, 
transportation tubes, and containers). 

Over the past decades, detailed knowledge has been accumulated in geology \cite{wald_form_1990,lorang_pebble_1990,szabo_reconstructing_2015,domokos_universality_2015}, 
physics
\cite{kun_transition_1999,katsuragi_scaling_2003,kadono_fragment_2005,
astrom_statistical_2006,carmona_fragmentation_2008,wang_dynamic_2011,PhysRevE.86.016113,Ye2019,astrom_statistical_2006,carmona_fragmentation_2008}, 
and engineering \cite{ma_wei_2018,andrews_frag_1999,arbiter-etal-69,
chau_fragmentation_2000,salman_classify_impact_2004} 
on single impact breakage phenomena, however, a comprehensive 
understanding of low velocity impact sequences responsible for 
the gradual mass reduction and  global rounding of solid particles is still lacking. {{In the physics literature
the existence of two distinct \emph{energy phases} (the \emph{damage phase} and the \emph{fragmentation phase}) has been established not only for brittle materials
\cite{carmona_fragmentation_2008,PhysRevE.86.016113,ma_wei_2018,myagkov_physicaa_2019}, 
and plastics spheres \cite{timar_new_2010}, but also for liquid droplets \cite{moukarzel_phase_2007}. Moreover, the same two energy phases have also been reported in the geophysics literature \cite{szabo_universal_2018} for the collisional attrition of sedimentary particles. In the latter context, global mechanical and geometric understanding of impact induced breakage would be  essential to decipher the information hidden in the size and shape of grains and pebbles \cite{wald_form_1990,lorang_pebble_1990,turcotte_fractals_1997, durian_what_2006,szabo_reconstructing_2015}.}} {{  
While research in geology and physics concentrated on single impact phenomena, mathematical research related to the proof of the Poincar\'e conjecture \cite{Gage_Hamilton_1986, Grayson_1987, Perelman_2003} led to the study of a class of nonlinear geometric partial differential equations (PDEs) called \emph{curvature-driven flows} which appear to be the adequate mean-field theory models for the global evolution of pebbles and other particles under a large number of low energy impacts \cite{Firey1974, Bloore1977}. One may target
global shape evolution of particles 
either by extending the physics literature about single breakage to multiple breakage processes or by relying on  
mean field PDE models.
Although the latter are invaluable tools to obtain qualitative insight, nevertheless, their application has not yet been rigorously justified: until now there existed no theory linking microscopic and macroscopic approaches, in particular, there were no clear physical criteria  established for the breakage process which would admit mean field PDEs as valid global approximations.

The latter, on the other hand, appear to be very useful as they make specific geometric predictions: global evolution starting from cuboid polyhedra  (serving as averaged models of natural fragments \cite{domokos_universality_2015,Domokos18178}) occurs in two \emph{geometric phases}: in the first (local rounding) phase vertices and edges become rounded but axis ratios hardly change while in the second (global rounding) phase  roundness remains almost constant while axis ratios increase \cite{Hamilton_Wornstones, Domokos2014}. These geometric phases have been identified  both in laboratory experiments and in numerical simulations of the PDEs \cite{Domokos2014, Priour_20} and this naturally led to the hypothesis that the geometric phases may also exist in a mechanical abrasion model.
In stark contrast to  geometric shape evolution of pebbles, \emph{no phases} can be distinguished in the evolution of mass \cite{Domokos2014} which appears to obey Sternberg's empirical law of exponential decay, approaching zero at infinite time \cite{sternberg}. }}


Here we present a thorough theoretical study of the phase structure of impact induced 
attrition processes { with the primary aim to establish a firm link between microscopic physical breakage models and macroscopic (mean field) geometric PDE models.} Based on realistic discrete element simulations  
of sequences of particle-wall collisions, we show that, 
depending on the impact velocity, the first (damage) energy phase may be clearly separated \emph{into two further energy phases}, so there exist three distinct energy phases: 
at sufficiently low velocities repeated impacts result in abrasion  of the body and lead to a finite asymptotic residual mass { (\emph{abrasion phase})}, however, above a threshold velocity 
a complete destruction is achieved within a finite number of repetitions { (\emph{cleavage phase})}. Instantaneous fragmentation
occurs above a second critical velocity where cracks span the entire body and the sample 
rapidly falls apart into a large number of small pieces { (\emph{fragmentation phase})}. The transitions between the abrasion, 
cleavage, and fragmentation phases occur at well-defined critical velocities analogous to continuous
phase transitions. 

{\gd  We establish the link between microscopic physical breakage models and mean-field PDEs in two steps.
First, the splitting of the earlier identified damage phase into the abrasion and cleavage phases
delineates the range of validity for the latter: the main feature of the now identified abrasion phase is that each impact removes only a small amount of (relative) mass. As PDE models are based on the limit where the
removed relative mass in each collision approaches zero, our study shows that  PDEs can be regarded as a mean field approximation of collision-induced attrition in the abrasion energy phase. Second, we identify one key feature of the PDE model in the microscopic simulation: we show that two \emph{geometric phases} earlier identified in the context of the PDE model clearly emerge inside the abrasion phase in the microscopic breakage model.}

Our finding is based on large scale computer simulations which revealed that the evolution of the mass and shape of the solid is governed by scaling laws in terms of the impact velocity. Most notably, { in the abrasion phase the shape evolution of the sample 
is described by a universal scaling form with a power law dependence on the impact velocity predicting infinite sample lifetime at some finite, asymptotic mass, the latter being determined by the energy threshold for the creation of cracks. In the special limit when this threshold approaches zero, our findings reproduce Sternberg's Law \cite{sternberg}, predicting exponential decay (and infinite lifetime) for sedimentary particles undergoing collisional abrasion in fluvial environments. In addition to verify Sternberg's Law for mass evolution, in the energetic abrasion phase we also confirmed the existence of the two earlier observed  \emph{geometric phases} \cite{Domokos2014, Hamilton_Wornstones}, thus our simulations serve as the first direct mechanical confirmation of curvature-driven PDEs as models of impact-driven abrasion processes.} In the cleavage
phase { we find that} the sample lifetime decreases as a power law of the impact velocity analogous to the  Basquin law\cite{basquin_exponential_1910,kun_universality_2008} of sub-critical fracture. 

\section{Single impacts: transition from damage to fragmentation}
To understand the evolution of solid bodies under repeated collisions with a hard wall,
first we focus on single impact events and quantify the resulting mass reduction.
We performed numerical measurements by means of computer simulations of a 
realistic discrete element model (DEM) of body-wall collisions in three dimensions (3D) 
\cite{PhysRevE.88.062207,PhysRevLett.112.065501,
pgergo_PhysRevE.90.062811,gergo_shearband_pre_2016,pal_granmat_2016} 
varying the impact velocity $v_0$ in a broad range. To represent freshly fractured rocks with sharp corners and edges in the initial state of shape evolution,
rectangular samples of mildly elongated cubic shape were created with the aspect ratio
$1:1.2:1.4$ of their shortest $c_0$, intermediate $b_0$, and longest $a_0$ sides. 
This choice is justified by our recent finding that the average shape of fragments is well approximated by a cube for a large diversity of fragmentation processes \cite{Domokos18178}.

In the model the sample is represented as a random packing,  consisting, on the average, of 12.000 spherical particles with a uniformly distributed diameter $d$ in a narrow interval 
$\Delta d$ around the average $\left<d\right>$ with $\Delta d/\left<d\right>=0.05$ 
\cite{pgergo_PhysRevE.90.062811}. Cohesive interaction 
is realized by beam elements which connect the particles along 
the edges of Delaunay triangles constructed from the initial particle 
positions \cite{PhysRevE.88.062207,PhysRevLett.112.065501,kun_roysoc_2019}. 
During the impact process in three dimensions (3D), the total deformation of a beam is calculated 
as the superposition of elongation, torsion, as well as, bending and 
shearing \cite{carmona_fragmentation_2008,wittel_intjfract_2008}. 
Cracks are formed when overstressed beams break according to a physical breaking rule. 
The breaking condition takes into account the stretching and shearing of particles contacts
\begin{eqnarray}
\left(\frac{\varepsilon_{ij}}{\varepsilon_{th}}\right)^2 + \frac{\max(\Theta_i, 
\Theta_j)}{\Theta_{th}} \geq 1,
\end{eqnarray}
where $\varepsilon_{ij}$ denotes the axial strain of the beam between particles $i$ 
and $j$, while 
$\Theta_i$, and $\Theta_j$ are the bending angles of the beam ends.
The parameters $\varepsilon_{th}$ and $\Theta_{th}$ control 
the relative importance of the two breaking modes 
\cite{kun_study_1996,carmona_fragmentation_2008,PhysRevE.86.016113,PhysRevLett.112.065501,
PhysRevE.88.062207}. 
The breaking criterion is evaluated after each iteration step and those beams which fulfill the condition are removed from the system.
In the model there is only structural disorder
present, i.e.\ the breaking thresholds are constant $\varepsilon_{th}=0.02$ and
$\Theta_{th}=3^o$, however, the physical properties
of beams such as length, cross section, and elastic moduli, are determined by the random 
particle packing. 
At the broken beams along the surface of the spheres cracks are
generated inside the solid and as a result of the successive beam breaking the
solid falls apart. The interaction of those particles which are not connected 
by beams, e.g.\ because the beam has been broken, is described 
by the Hertz contact law \cite{farhang_book_2011}. 
The time evolution of the impacting body is generated by solving 
the equation of motion of all the particles.
Parameters of the model were set in such a way that our DEM provides a consistent qualitative and in certain cases quantitative description of the mechanical and fracture properties of the broad class of heterogeneous brittle materials which are abundant in our geological environment 
(see Table \ref{tab:table_1}). 
\begin{figure}
\begin{center}
\includegraphics[bbllx=50,bblly=40,bburx=660,bbury=470,scale=0.39]{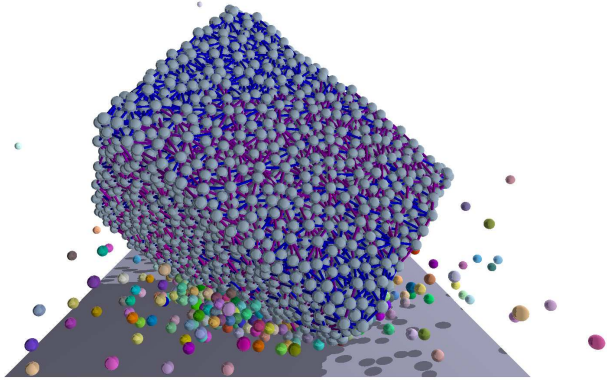}
  \caption{A snapshot of the time evolution of the first impact of a sample with a hard wall. 
  The initially angular body hits the 
  wall close to one of its corners. Most of the fragments are single particles flying at a 
  high speed. Colors are randomly assigned to the fragments. The intact cohesive contacts 
  are represented by lines connecting the spherical particles.
   \label{fig:snapshot}}
\end{center}
\end{figure}
The model has been successfully applied 
before to investigate fracture and fragmentation under various types 
of loading conditions \cite{PhysRevLett.112.065501,PhysRevE.88.062207,
pgergo_PhysRevE.90.062811,gergo_shearband_pre_2016}. 

Initially, the discretized sample was placed close to a planar wall with a random orientation
chosen uniformly on the sphere
and the impact was initiated by assigning identical velocity $v_0$ 
perpendicular to the wall to all particles of the solid . 
As the body moved, it  got into contact with the wall and 
deformed  which could result in cracking and fragment formation. 
The impact lasted until complete rebound 
was  achieved where all particles separated from the wall. In the final state of the process, particles connected by the surviving cohesive elements were identified as fragments.
A snapshot of the impact process is presented in Fig.\ \ref{fig:snapshot}.

\begin{table}
\caption{{\bf Parameters of the discrete element model.} \label{tab:table_1}} 
\begin{tabular}{l}
   \hline
   \T
   {\bf Beams:}\\
   \begin{tabular}{lllll}
     & longitudinal stiffness~~~~~~~~~~~~~~~~~ & $E^b$~~~ & 6~~~~~~~~~~ & GPa\\
     & strain threshold & $\varepsilon_{th}$ & 0.02 & -\\ 
     & bending threshold & $\theta_{th}$ & 3 & $degree$\\
   \end{tabular} \\
   {\bf Particles:}\\
   \begin{tabular}{lllll}
     & stiffness~~~~~~~~~~~~~~~~~ & $E^p$~~~~~~~ &3~~~~~~~~~~ & GPa\\
     & Average diameter& $\left<d\right>$ & 0.5 & mm\\
     & density & $\rho$ & 3000 & kg/m$^3$\\ 
   \end{tabular} \\
   {\bf Hard wall:}\\
   \begin{tabular}{lllll}
     & stiffness~~~~~~~~~~~~~~~~~ & $E^w$~~~~~~~ &70~~~~~~~~~ & GPa\\
   \end{tabular} \\
   {\bf Interaction:}\\
   \begin{tabular}{lllll}
     & friction coefficient~~~~ & $\mu$~~~~~~~~~ & 1~~~~~~~~~~ & -\\
     & damping coefficient (normal) & $\gamma_n$ & 0.25& s$^{-1}$\\ 
     & friction coefficient (tangential) & $\gamma_t$ & 0.05 & s$^{-1}$\\ 
   \end{tabular} \\
   {\bf System:}\\
   \begin{tabular}{lllll}
     & time increment~~~~~~~~ & $\Delta t$~~~~~~~ & 1e-7~~~~~~ & s\\
     & average number of particles & $N^p$ & 12000 & -\\ 
     & average number of beams & $N^b$ & 105000 & -\\ 
     & solid fraction &  & 0.65 & -\\ 
   \end{tabular} \\ 
   \\
   {\bf Macroscopic properties (DEM):}\\
   \hline
   \begin{tabular}{lllll}
     \T
     & system stiffness~~~~~~~~ & $E$~~~~~~~~ & $7.4\pm 0.5$ & GPa\\
     & Poisson's ratio & $\nu$ & 0.2 & -\\ 
     & system strength & $\sigma_c$ & 110 & MPa\\ 
   \end{tabular} \\
   \hline
 \end{tabular}
\end{table}

Simulations revealed that  for sufficiently low impact velocities $v_0< v_a$  the sample solely underwent deformation around the impact site and rebounded
elastically without suffering 
any damage. Cracks 
first occurred when $v_0$ 
surpassed a threshold velocity 
$v_a$ determined by the strength of the internal cohesive elements of the material. 
In this low velocity range, deformation and crack formation is restricted to the 
vicinity of the contact zone, while 
for high impact velocities 
cracks can span the entire sample giving rise to 
rapid breakup.
\begin{figure}
\begin{center}
\includegraphics[bbllx=30,bblly=30,bburx=380,bbury=330,scale=0.7]{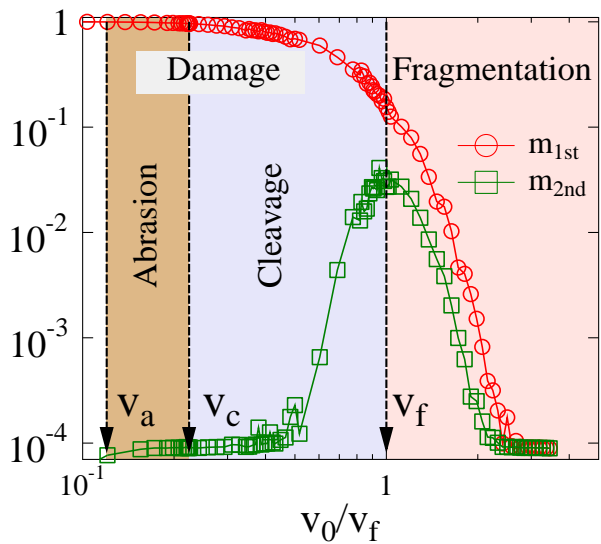}
  \caption{The three energy phases. Mass of the largest $m_{1st}$ and second largest $m_{2nd}$ fragments obtained 
  after a single impact as function of the impact velocity $v_0$. 
In the regime $v_0<v_a$ of low velocities no cracking occurs and the impactor elastically rebounds from the wall. In the abrasion phase $v_a<v_0<v_c$ small fragments are 
removed from the body by chipping. To achieve complete breakup in a single impact, $v_0$ has to exceed
the critical fragmentation  velocity $v_f$. 
In the intermediate velocity range
$v_c<v_0<v_f$ of cleavage, cracks penetrate deeper inside splitting larger pieces from the body.  
The critical velocity $v_c$ of cleavage is the threshold velocity above which the asymptotic remaining mass tends to zero in repeated collisions.
Horizontal axis shows on  logarithmic scale the impact velocity $v_0$ normalized by critical fragmentation velocity $v_f$.
For the model solid we found the  (non-dimensional) ratio of
threshold velocities to be
$v_a/v_f=0.124$.
and 
$v_c/v_f=0.224$.
   \label{fig:max_1_2}}
\end{center}
\end{figure}
To give a quantitative characterization of the final outcome and the degree of destruction 
caused by impacts, we determined the average masses $M_{1st}$ and $M_{2nd}$ of the largest and second largest
fragments, respectively. After normalizing these values by the total mass $M_0$ we plotted $m_{1st}=\left<M_{1st}/M_0\right>, m_{2nd}=\left<M_{2nd}/M_0\right>$  as function of the 
impact velocity $v_0$.
It can be observed in Fig.\ \ref{fig:max_1_2}
that at low impact velocities  we have $m_{2nd}\ll m_{1st}$, i.e. the second largest fragment is orders of 
magnitude smaller than the largest one, showing that only small 
pieces are removed from the body around the impact site. This 
is characteristic for the \emph{damage energy phase}.
Fragmentation is achieved when the second largest piece becomes 
comparable to the largest one, which first occurs at the maximum of $m_{2nd}$ defining the critical velocity $v_f$ of fragmentation. Beyond the \emph{critical  fragmentation velocity} $v_f$  both $m_{1st}$ and $m_{2nd}$ decrease monotonically.
Figure \ref{fig:max_1_2}, 
illustrating the phase diagram of impact induced attrition processes,
shows that, depending on the velocity, single impacts  
give rise either to damage 
or fragmentation of the sample with a sharp transition at the critical velocity $v_f$.  
The damage - fragmentation transition has already been studied 
in experiments and computer simulations  of impacting spherical samples 
against a hard wall, using heterogeneous brittle materials
\cite{carmona_fragmentation_2008,PhysRevE.86.016113,ma_wei_2018,myagkov_physicaa_2019}, 
plastics spheres \cite{timar_new_2010}, and liquid droplets \cite{moukarzel_phase_2007}.  In these studies,
the same qualitative behavior was obtained for the largest fragment masses $m_{1st}$, $m_{2nd}$ 
as in Fig.\ \ref{fig:max_1_2}, which implies that
 the overall outcome of the process in the high velocity range is entirely controlled by the impact velocity and its critical value $v_f$, whereas  neither the sample's shape  nor 
materials' features have any relevant effect.
The detailed analysis of the 
mass distribution of fragments revealed that the observed universality is caused by the underlying 
continuous phase transition from damage to fragmentation as the impact velocity is varied
\cite{kun_transition_1999,kadono_fragment_1997,moukarzel_phase_2007}. The identification of the 
known damage and fragmentation phases also serves as a verification of our model.

\section{Repeated impacts and the two sub-phases of damage: abrasion and cleavage}
In the previous subsection, confirming earlier results, we established for single impact phenomena the existence of the two main energy phases. Now we will show that,  if we consider not just a single impact but impact \emph{sequences}, the damage phase can be subdivided into two narrower energy phases: abrasion and cleavage.
The damage phase, characterized by $v_0\ll v_f$ is often observed in natural and industrial processes at lower energy levels. Under such conditions, the large residue of the sample typically undergoes repeated 
collisions which give rise to a complex evolution of its size and shape.
In the following we extend the global phase diagram of Fig.\ \ref{fig:max_1_2} 
refining the structure of the damage phase by characterizing qualitatively different 
evolution histories of residues under \emph{ repeated} sub-critical impacts.

To simulate sequences of particle-wall collisions, 
in the final state of an impact event we identified the largest fragment as the residue of the body. It was then replaced by its counterpart cut out of the initial sample, to ensure an undeformed, relaxed initial state for the subsequent collision.
The residue was randomly rotated in three dimensions and was impacted against the wall with the same impact velocity $v_0<v_f$ as before.
The above procedure was repeated up to $N_{max}=400$ times, or until complete destruction of the body, 
at $\approx 60$ different
impact velocities, respectively. For each sequence, $120$ different
initial samples were used, while in subsequent impacts the residues were randomly rotated  by uniformly choosing a direction on the sphere.
These calculations revealed an astonishingly rich phase structure of the sub-critical $v_0<v_f$ regime.

\begin{figure}
\begin{center}
\includegraphics[bbllx=20,bblly=20,bburx=730,bbury=330,scale=0.35]{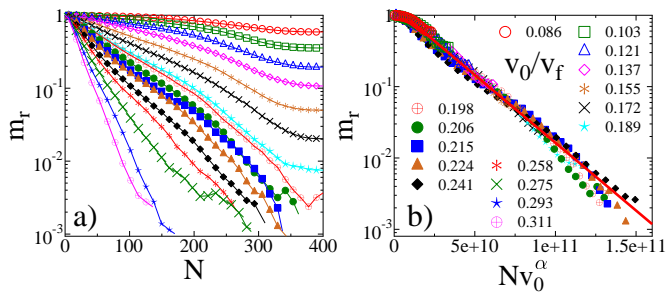}
  \caption{$(a)$ Average mass of the residue $m_r$ as a function of the 
impact number $N$ for several impact velocities $v_0$ below the fragmentation
critical point $v_f$. Panel $(b)$ shows that by
rescaling the horizontal axis, curves of different $v_0$ values
can be collapsed onto one single master curve. 
Note that data of the lowest impact velocities (highest remaining mass) are in the upper left corner at the start of the master curve.
Straight line represents the exponential 
form of the scaling function $\widetilde{m}_r(x)$ of Eq.\ (\ref{eq:scalmass}).
The legend for $(a)$ and $(b)$ is given in $(b)$.
   \label{fig:residue}}
\end{center}
\end{figure}

To quantify the gradual mass reduction during the collision sequence, 
Figure \ref{fig:residue}$(a)$ presents the average $m_r=\left<M_r/M_0\right>$ 
of the residual mass $M_r$ normalized by the initial mass of the  sample $M_0$,
as a function of the impact number $N$ for several values of $v_0$.  (We remark that for a single impact event we have $m_r \equiv m_{1st}$.)
At very low velocities $v_0\ll v_f$, a single impact always gives 
rise only to a few fragments which are typically single spheres, 
i.e.\ powder in the model. 
As a consequence, in Fig.\ \ref{fig:residue}$(a)$ the residual mass $m_r$ gradually 
decreases with increasing impact number $N$, however, mass reduction gets limited 
for high $N$ values and a finite asymptotic residual 
mass emerges $m_r\to m_r^a$  as $N \to \infty$.
The reason is that due to the decreasing mass $M_r$, the kinetic energy 
$ E_0 = \frac{1}{2}M_rv_0^2$,
imparted to the sample decreases, since the impact velocity $v_0$ is fixed.
Consequently, beyond a certain impact number (i.e.\ below a certain value of $M_r$),
the emerging deformation is not sufficient to induce further cracking.

Since only small pieces are removed in single impacts, we term this velocity 
regime as the \emph{abrasion phase} of the system characterized by the existence of a finite asymptotic 
residual mass $m_r^a>0$. 
It can be observed in Fig.\ \ref{fig:residue}$(a)$ that the value of $m_r^a$ decreases 
with increasing impact velocity $v_0$. The value of $m_r^a$  depends also  on the energy threshold for the creation of cracks, i.e.\ on the strength of cohesive contacts. In the limit when this threshold approaches zero, our findings reproduce Sternberg's Law \cite{sternberg}, predicting exponential decay  to zero mass and infinite lifetime for sedimentary particles undergoing collisional abrasion in fluvial environments.

When $v_0$ gets sufficiently high, the functional form of $m_r(N)$ qualitatively changes:  
the  mass of the residue sets to a rapid decrease with $N$, and repeated impacts 
give rise to a complete destruction of the sample within a finite number of repetitions. 
This behavior is characterized by impact velocities in the range  $v_c  < v_0 < v_f$  and we call this interval the \emph{cleavage phase} of the impact sequence. The critical velocity  $v_c$ of cleavage is defined as the threshold velocity above which the asymptotic residual mass is zero even at finite energy threshold for the creation of cracks.  
In our discrete element model, a complete destruction
of the sample is reached when the largest fragment comprises solely a single particle of the  discretization. For real materials this state is realized when the residual size approaches  a characteristic length scale of the meso-structure, e.g.\ grain size of materials.

The transition from abrasion to cleavage at the critical velocity $v_c$ is driven by the changing mechanism of cracking.
In the abrasion phase the dominating mechanism of mass removal is chipping, i.e.\ crack formation 
parallel to the contact surface with the wall, which leads to the formation of tiny fragments
\cite{subero_breakage_2001,ghadiri_impact_2002}. 
However, in the case of cleavage, cracks penetrate the solid to significantly 
deeper regions so that a combination of contact damage and fracture occurs,
giving rise to coarser products as well. Additionally, the elastic 
waves generated by the collision give rise to the gradual accumulation of damage  inside the residue which, in turn, can result in fatigue crack growth as the impact sequence proceeds \cite{herrmann_statistical_1990}.

Our results demonstrate that above the threshold velocity of micro-cracking $v_a$,  impact attrition phenomena have additionally two well-defined critical impact velocities $v_c$ and $v_f$, which separate 
the three phases of abrasion, cleavage, and fragmentation with distinct 
qualitative behaviors (see  the phase diagram of Fig.\ \ref{fig:max_1_2}). 
For our model solid, the threshold velocities of abrasion and cleavage are $v_a/v_f=0.124\pm 0.004$ and 
$v_c/v_f=0.224\pm 0.005$, with respect to the fragmentation critical velocity $v_f$.

\subsection{Sternberg's law and Basquin's law}
Figure \ref{fig:residue}$(b)$ demonstrates that rescaling the impact 
number $N$ with a proper power $\alpha$ of $v_0$,  curves belonging to different
impact velocities $v_0$ can be collapsed on the top of each other, yielding the scaling form
\begin{equation}
m_r(N,v_0) = \widetilde{m}_r(Nv_0^{\alpha}),
\label{eq:scalmass}
\end{equation}
where the scaling function $\widetilde{m}_r(x)$ can be approximated by an 
exponential $\widetilde{m}_r(x) \sim \exp{(-x)}$ (see Fig.\ \ref{fig:residue}$(b)$), reproducing the time evolution predicted by Sternberg's law \cite{sternberg}. 
Best collapse is achieved in Fig.\ \ref{fig:residue}$(b)$ with the exponent $\alpha=2.1\pm 0.15$. 

It follows from the scaling analysis that increasing 
the impact velocity $v_0$ the characteristic impact number $N_c$ of the time evolution 
decreases as a power law 
\begin{equation}
N_c\sim v_0^{-\alpha}. 
\label{eq:n_c}
\end{equation}
The scaling law Eq.\ (\ref{eq:n_c}) holds in both the abrasion and cleavage phases $v_a<v_0<v_f$ of impact attrition. For cleavage, the characteristic impact number $N_c$  can be interpreted as the lifetime of the sample.
Since the peak stress, emerging at the contact zone during impact, increases as a power 
of the impact velocity $v_0$ \cite{johnson_contact_book}, it follows 
that the expression  (\ref{eq:n_c}) of residual lifetime is analogous to the Basquin law 
of sub-critical fracture phenomena
\cite{basquin_exponential_1910,suresh_fatigue_1998,sornette_physical_1992,kun_universality_2008,vieira_subcritical_2008}. The Basquin law of fatigue life is a fundamental law of sub-critical 
fracture. It expresses that under a constant or varying sub-critical load, where the stress 
amplitude falls below the fracture strength of materials, failure occurs
in a finite time which decreases as a power law of the externally applied
stress amplitude \cite{basquin_exponential_1910}. 
Our results demonstrate that the Basquin law holds also for sub-critical 
impact phenomena.

\begin{figure}
\begin{center}
\includegraphics[bbllx=40,bblly=30,bburx=380,bbury=330,scale=0.70]{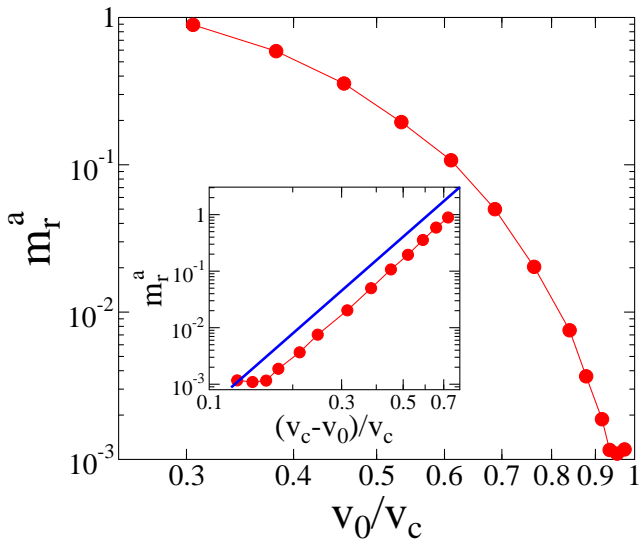}
\caption{Asymptotic mass $m_r^a$ of the residue as a function of the impact velocity in the abrasion
  phase $v_0<v_c$. 
  Inset: the mass values of the main panel are re-plotted as a function of the relative distance
  from the critical point $v_c$. The straight line represents a power law of exponent 
  $\beta=4.2$.
   \label{fig:resmass_crit}}
\end{center}
\end{figure}
In the abrasion phase 
$N_c$ characterizes the rate of convergence to the asymptotic residual mass $m_r^a$. 
Additionally, the impact velocity also determines the value of $m_r^a$, which tends to zero 
when approaching the critical point $v_c$ from below, see Fig.\ \ref{fig:resmass_crit} which also shows
(inset)
that the convergence to zero is well described by a power law as a function of the distance from the critical point
\begin{equation}
  m_r^a \sim (v_c-v_0)^{\beta}, \qquad \qquad \mbox{for} \qquad v_0\leq v_c.
\end{equation}
For the exponent we obtained $\beta=4.2\pm 0.2$ by fitting of data.
Since in the the cleavage phase we have  $m_r^a=0$, whereas in 
the abrasion phase we have $m_r^a>0$, 
$m_r^a$ can be considered as the order parameter 
of the abrasion-cleavage phase transition, and $\beta$ is the order parameter exponent
of the transition.

\section{Shape evolution: geometric phases inside the abrasion energy phase}

\subsection{Mean field models}
In case of polyhedral initial samples, in the abrasion phase we  expect that at the beginning of the impact sequence 
sharp corners and edges are gradually removed, giving rise
to an evolution towards an asymptotic rounded shape. 
In the cleavage phase, due to the breaking of coarser pieces, this evolution 
is more erratic and eventually results in an ultimate destruction.
Due to the small size of fragments, we expect that geometric aspects of the abrasion phase may be well reflected in the solutions of averaged, mean field geometric PDE models of attrition \cite{Firey1974, Bloore1977}. The simplest, two-dimensional version of these PDE models may be written as 
\begin{equation}\label{eq:curv}
    V=c\kappa,
\end{equation}
where $V$ denotes the speed by which a surface points moves inward along the surface normal, $\kappa$ is the scalar curvature and  equation (\ref{eq:curv}) is often referred to \cite{Gage_Hamilton_1986, Grayson_1987} as the \emph{curve shortening flow} or as the geometric heat equation. The constant $c$ can be regarded as scaling of time and plays no role if evolution is plotted as a function of the normalized residual mass $m_r$. (We remark that equation (\ref{eq:curv}) is written in a compact, invariant notation, details about this and other notations are given in Section 1 of the Supplemental Material 
\cite{suppl_prx}.)
Next we will show that these expectations are well founded and PDE models serve indeed as good approximations of impact-induced attrition processes, however, only in the abrasion phase.

\subsection{Shape descriptors}
To give a quantitative characterization of the rounding process, we picked three dimensionless descriptors 
of the overall shape of the residue (thus neglecting morphological details of its surface) which not only provide efficient monitoring of the geometric evolution but also admit meaningful comparison with earlier results: axis ratios, circularity (isoperimetric ratio) and intact surface ratio.   
\begin{enumerate}
\item
\emph{Axis ratios} $c/a$ and $b/a$ are traditional geological descriptors \cite{Domokos2014} characterizing the shape of the residue \cite{pen_dependence_2013,domokos_universality_2015} where $a>b>c$ 
refer to the axes of the bounding box of the residue, aligned with the edges of the initial (cuboid) sample. 
\item
\emph{Isoperimetric ratio} or circularity of a planar object is given as $R=4\pi A/P^2$, where $A,P$
refer to area and perimeter, respectively. It has been observed \cite{szabo_universal_2018} that circularity of the 
largest projection of sedimentary particles shows universal features in fluvial abrasion and its evolution
is entirely determined by the mass loss during impact induced attrition processes. In our DEM, $A$ and $P$ of the residue were obtained as the area and perimeter of the convex hull of the point cloud of the largest projection of the spherical particles of the relaxed body. For more details on shape descriptors see Section 3 of the Supplemental Material \cite{suppl_prx}.

\item \emph{Intact surface ratio} $S/S_0$, expressing the intact fraction of the initial surface, was selected following an idea of Richard Hamilton \cite{Hamilton_Wornstones} who, in one of the papers dedicated to the study of curvature-driven flows (leading ultimately towards to his seminal contribution to the proof of the Poincar\'e - conjecture) describes a curious nonlinear phenomenon about intact surface ratio in the Gauss curvature flow which is the 3D version of (\ref{eq:curv}): he predicted that $S/S_0$ will drop to zero after a finite time, marking the end of the first geometric phase for cuboids. (For more details see Section 4 of the Supplemental Material \cite{suppl_prx}.)  In the initial state of DEM samples $S_0$ is determined as the number of particles covering the external body surface, then the surviving intact surface $S$ is obtained by tracing the particles removed from the initial 
surface $S_0$ in subsequent impacts.
\end{enumerate}
{\gd  The evolution of axis ratios $c/a$ and $b/a$  and the evolution of the isoperimetric ratio $R$ has been computed in the PDE model \cite{Domokos2014} for the very same cuboid initial conditions as in our DEM study. For the evolution of intact surface ratio $S/S_0$
in the PDE model we have an analytical result \cite{Hamilton_Wornstones}.  We will now establish the link  between PDE models and microscopic computations by comparing these evolutions. The most striking qualitative feature of the PDE model is the spontaneous emergence of two \emph{geometric phases} and our computations reveal that these phases are perfectly captured in the microscopic DEM approach. To make the comparison between plots for shape descriptors meaningful,  next we seek the corresponding scaling laws.}

 \begin{figure}
\begin{center}
\includegraphics[bbllx=30,bblly=50,bburx=750,bbury=650,scale=0.33]{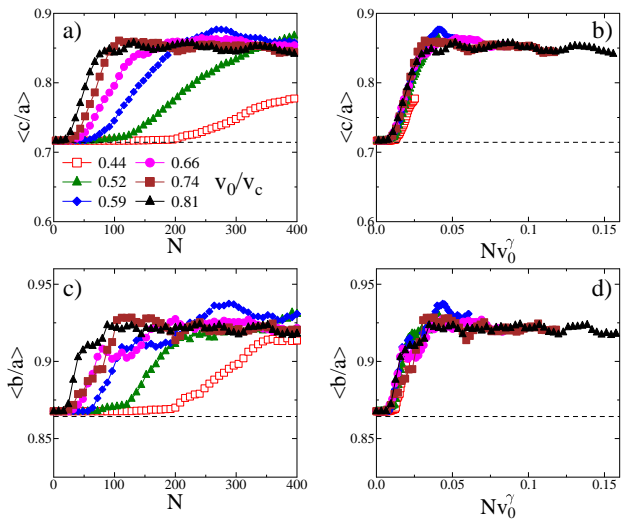}
  \caption{  Average side length ratios $\left<c/a\right>$ $(a)$ and $\left<b/a\right>$ $(c)$ 
  of the bounding box of the residues as function of the impact number $N$ for different  
  impact velocities inside the abrasion energy phase $v_0<v_c$. 
  The horizontal dashed lines represent the initial values $\left<c_0/a_0\right>=1/1.4$ and 
  $\left<b_0/a_0\right>=1.2/1.4$.
  By rescaling the impact number $N$ in $(b)$ and $(d)$ by an appropriate power $\gamma$ 
  of the impact velocity $v_0$, the curves of different $v_0$ of $(a)$ and $(c)$ can be collapsed on master curves. Best collapse is achieved using the same exponent  $\gamma=3$ in $(b)$ and $(d)$.
   \label{fig:bounding_box}}
\end{center}
\end{figure}
\subsection{Scaling laws}
Increasing $v_0$ accelerates mass removal and thus shape evolution. 
Figures \ref{fig:bounding_box}$(a,c)$ demonstrate that both
axis ratios $\left<c/a\right>(N,v_0)$, $\left<b/a\right>(N,v_0)$ remain initially constant,
display sudden growth between the characteristic impact numbers $N_{r}$ and $N_s$ and subsequently
saturate. Both the overall shape of these functions and their saturation values remain the same in the entire abrasion phase, however both $N_{r}$ and $N_s$ decrease with increasing $v_0$. We found that 
rescaling these curves 
with $v_0^{\gamma}$, they collapse onto master curves (see Fig.\ \ref{fig:bounding_box}$(b,d)$) implying the scaling structure 
\begin{eqnarray}
\left<c/a\right>(N,v_0) &=& \Phi(Nv_0^{\gamma}), \\
\left<b/a\right>(N,v_0) &=& \Psi(Nv_0^{\gamma}),
\label{eq:ratio_scaling}
\end{eqnarray}
where $\Phi(x)$ and $\Psi(x)$ denote the scaling functions. This also implies that
$N_{r}$ and $N_s$
both have the same power law dependence
\begin{equation}
 N_{r} \approx A v_0^{-\gamma}, \qquad \qquad N_{s} \approx B v_0^{-\gamma},
 \label{eq:nce_ns_scaling}
\end{equation}
where the exponent $\gamma$ was obtained numerically $\gamma=3.0\pm 0.07$. The saturation values 
$\left<c/a\right>\approx 0.865$ and $\left<b/a\right>\approx 0.925$ show that the asymptotic stable shape 
of the object is slightly anisotropic which may be a consequence of the finite number of the non-breakable discrete elements in the simulation.
Our simulations revealed that under the condition 
of isotropic impacts, the origin of the universal 
scaling forms is that the shape of the evolving object is controlled by the total relative mass
$\mu(N)=1-m_r$ lost in $N$ repeated collisions.
Recently, it has been suggested \cite{szabo_universal_2018} that  $\mu(N)$
is also controlling the evolution of the circularity $R$, so
henceforth we use this representation for all shape descriptors.

\subsection{Geometric phases}

\begin{figure}[h]
\begin{center}
\epsfig{file=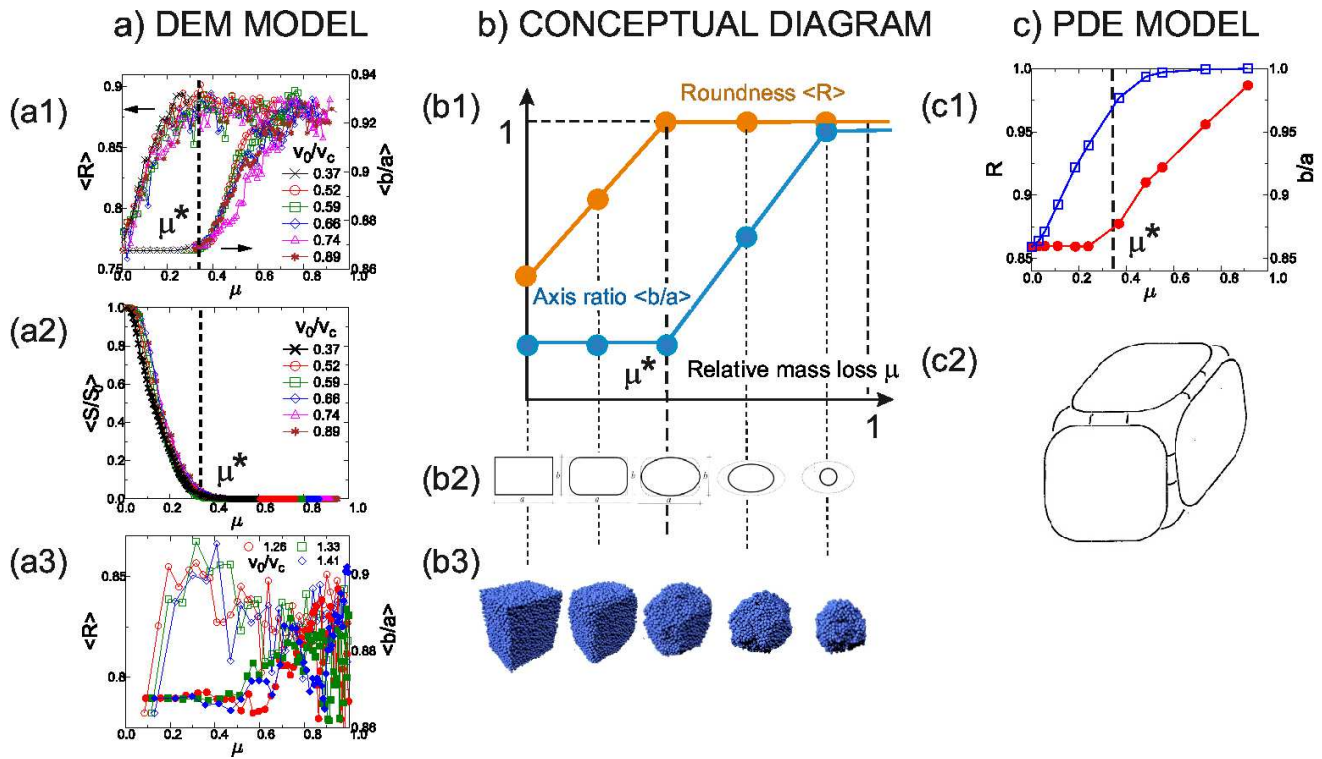,bbllx=0,bblly=0,bburx=1120,bbury=650,width=17.5cm}
  \caption{Geometric shape evolution as a function of the relative abraded mass $\mu$. (a) DEM simulations: (a1) Abrasion energy phase: evolution of the average circularity $\left<R\right>$ and axis ratio $\left<b/a\right>$.  Observe two geometric phases: phase 1 with approximately constant $\left<b/a\right>$ followed by phase 2 with approximately constant $\left<R\right>$. Transition at $\mu^* \approx 0.34$. (a2) Abrasion energy phase: evolution of intact surface ratio $S/S_0$.  Transition between phases at $\mu^* \approx 0.34$. (a3) Cleavage phase: evolution of the average circularity $\left<R\right>$ (open symbols) and axis ratio $\left<b/a\right>$ (filled symbols).  Observe absence of smooth evolution. (b1) Schematic, bilinear approximation of two-phase geometric evolution
  of axis ratios and roundness. (b2) Schematic side view of abrading cuboids, projected along the shortest ($c$) axis. (b3) Snapshots of DEM simulations.  (c) PDE model results: (c1) Evolution of circularity $R$ and axis ratio $b/a$ \cite{Domokos2014}. Observe two geometric phases: phase 1 with approximately constant $b/a$ followed by phase 2 with approximately constant $R$. (c2) Hand-drawn sketch by R. Hamilton \cite{Hamilton_Wornstones} predicting phase 1 characterized by nonzero intact surface ratio.   
   \label{fig:0}}
\end{center}
\end{figure}

The PDE model (\ref{eq:curv}) predicts for the evolution of cuboid blocs with moderate initial axis ratios the emergence of two geometric phases: in phase 1 axis ratios $c/a, b/a$ remain approximately constant while roundness increases steeply and saturates close to 1. In phase 2 the opposite happens: axis ratios increase steeply and saturate close to 1 while roundness remains constant. The conceptual plot of this evolution (as a function of the relative abraded mass $\mu=1-m_r$) is shown in Figure \ref{fig:0}(b1), accompanied by conceptual contours of the specimen, projected along the shortest ($c$) axis (b2)  and representative snapshots of DEM simulations (b3). Figure \ref{fig:0}(c1) shows the same plot for $b/a$ and $R$, obtained from the numerical computation \cite{Domokos2014} of the PDE (\ref{eq:curv}). Figure \ref{fig:0}(c2) presents the hand-drawn sketch of Hamilton \cite{Hamilton_Wornstones} of his analytical result on the same PDE: intact surface area $S/S_0$ survives for a finite time and this marks geometric phase 1.

In Figure \ref{fig:0}(a) we compare the DEM computations to the aforementioned analytical predictions. In Figure \ref{fig:0}(a1) we show evolutions of the average axis ratio $\left<b/a\right>$ and roundness $\left<R\right>$  in the abrasion energy phase
$v_a< v_0<v_c$. Note that curves of different impact velocities all fall on the top of each other in agreement with the scaling collapse predicted in the previous section. It is apparent that we have good qualitative agreement with Figure \ref{fig:0}(c1): $\left<b/a\right>$
remains constant  at the initial value $\left<b/a\right>=1.2/1.4$ until $\mu^*\approx 0.34$ while $\left<R\right>$ increases sharply
and the opposite can be observed for $\mu>0.34$. 
Based on this observations 
we can clearly record the presence of the two geometric phases  in the abrasion energy phase
for the evolutions of the axis ratios and the roundness.
\onecolumngrid

\twocolumngrid

 In Figure \ref{fig:0}(a2) we show the evolution of the intact surface ratio $S/S_0$, also in the abrasion energy phase
$v_a< v_0<v_c$.  We can observe that this shape descriptor drops to zero at the same relative abraded mass value ($\mu^* \approx 0.34$) which separates the two phases for the evolution of axis ratios and roundness. This is in agreement with the prediction of Hamilton \cite{Hamilton_Wornstones} who claimed that intact surface area will survive for a finite time. It is easy to see that as long as intact surface area exists, the corresponding axis ratio of the cuboid (computed from the bounding box) will remain constant so here again we see a perfect match between the DEM computations and the prediction based on the PDE. The transition point $\mu^*$ between the two geometric phases in the microscopic DEM and macroscopic mean field PDE descriptions of shape evolution have a very good agreement.  

This confirms our claim that in the abrasion energy phase $v_a< v_0<v_c$ the PDE model offers adequate description of the shape evolution. In sharp contrast, Figure \ref{fig:0}(a3) illustrates the evolution of the axis ratio $b/a$ and roundness $R$ in the cleavage energy phase $v_c< v_0<v_f$, both displaying a non-smooth behavior:  here we do not expect any mean-field PDE model to provide an adequate description.

\section{Discussion}
Impact induced attrition processes cover a broad variety of
phenomena ranging from the gentle removal of fine powder from the surface of rock pieces 
by low velocity impacts
to the immediate disruption of objects in energetic collisions. 
Understanding gradual mass removal due to a sequence of impact events is crucial  in sedimentology since pebbles
can be considered as witnesses of the geological conditions of their creation.
Universal scaling laws of lifetime, size, and shape of evolving particles 
are indispensable to decode the information imprinted in pebbles \cite{wald_form_1990,lorang_pebble_1990,turcotte_fractals_1997, durian_what_2006,szabo_reconstructing_2015}. 
In the initial state of this evolution process freshly fragmented rocks are generated \cite{domokos_universality_2015}
by dynamic breakup of rock masses due to high velocity impacts.
While the theory of single impact of solid particles with a hard wall is well understood at
the level of particle-based models, impact sequences have been so far only modeled
by mean field theory which necessarily included gross simplifications of the breaking process.
Here we offered the first link between particle-based models and mean field theory for collision sequences.
 
The main methodological novelty of our study is that we use the discrete element method 
to realistically simulate the entire physical process of all the individual impacts of long 
sequences without any additional assumption. 
Although at high computational costs  (by simulating $\approx 5 \times 10^{5}$
collisions with samples consisting of $\approx 12.000$ discrete elements), this approach enabled us to unveil 
the rich phase structure of impact induced attrition processes.
Based on experimental observations, a descriptive classification of single impact 
breakage has been proposed in \ \cite{salman_classify_impact_2004}, where low, intermediate,
and high velocity ranges were distinguished according to the amount and structure of the resulting 
damage of the body. Here we demonstrated that in multiple impact processes
these regimes are separated by universal phase transitions.
In addition to the already known damage and fragmentation phases (separated
by the critical impact velocity $v_f$) we identified the abrasion and cleavage phases \emph{inside} the damage phase
(separated by the critical velocity $v_c$).
Abrasion results in finite asymptotic mass (analogous to Sternberg's Law \cite{sternberg})
while  cleavage results in a complete
destruction after a finite number of impacts, with sample lifetime decreasing 
as a power law of the impact velocity (analogously  to  Basquin's law). 

By identifying the abrasion energy phase we were able to provide the link between microscopic, 
particle-based models and mean-field curvature-driven equations. We showed that the latter can be regarded as adequate approximations of the former, however, only in the abrasion phase.
Our simulations revealed an astonishing universality
of the evolution of rounding of the residue. Both the axis ratios and the circularity of the largest projection
proved to be entirely determined by the attrition mass:  evolutions at different impact velocities $v_0$ can
be collapsed onto a single curve by rescaling  the number of impacts with a proper power $\alpha$ 
(also called the \emph{lifetime exponent}) of $v_0$.
This universality confirmed earlier conjectures  and observations \cite{Domokos2014, szabo_universal_2018} on the existence of two geometric phases  and also helped to identify a scaling law  of the dynamics: the characteristic event number of the onset of shrinking of 
initially angular objects proved to decrease as a power law of the impact velocity. 
We were also able to verify a curious effect of geometric nonlinearity, first predicted by Hamilton \cite{Hamilton_Wornstones}: in case of polyhedral initial shapes, a finite amount of the initial surface area survived abrasion for a finite amount of time.

Our findings also fit into the broader picture of efforts to approximate PDE models by microscopic, particle-based simulations. In the context of curvature-driven surface evolution, closest to our current topic, Monte-Carlo simulations of the Kardar-Parisi-Zhang (KPZ) equation proved to be a powerful tool to understand the global dynamics \cite{odor_2008_universality_book, odor_kpz_pre_2010}. However, in contrast to our approach, discrete KPZ models do not use a mechanics-based DEM Kernel and most often they are aimed at surface growth in an orthogonal $[xyz]$ frame.

It is important to emphasize that the excellent qualitative and quantitative agreement (e.g.\ for the transition point between the two geometric phases) of the microscopic DEM and macroscopic PDE descriptions of shape evolution were obtained without any parameter tuning of DEM simulations. This confirms the high degree of robustness of the results for the broad class of heterogeneous brittle materials. For the initial state of shape evolution we considered mildly anisotropic cuboids, since it has proven to be the generic average shape of freshly fractured rocks \cite{Domokos18178}. Cuboids with other axis ratios would only change the time scale of shape evolution and shift the transition point $\mu^*$ between the geometric phases. Inside the energy phases of abrasion and cleavage, the temporal evolution of mass and shape is controlled by the impact velocity which we could cast into scaling laws. The value of the scaling exponent of lifetime (cleavage) $\alpha$ falls close to 2, while the exponent $\gamma$ controlling the shape evolution (abrasion) has a higher value $\gamma\approx 3$. Based on fracture mechanics, approximate analytical expressions have been derived for the threshold velocities of the onset of abrasion $v_a$ and fragmentation $v_f$ \cite{ghadiri_impact_2002}. 
These calculations showed that the critical velocities separating the energy phases of impact attrition phenomena depend on material properties as well as on the mass and linear extension of the sample \cite{ghadiri_impact_2002}. Based on the analogy to continuous phase transitions, we conjecture that the critical exponents $\alpha$, $\beta$, and $\gamma$ are universal, they depend neither on mechanical, nor on geometrical features of the system.

\section*{Acknowledgments}
The authors thank Andr\'as Sipos for his invaluable help with computing Figure \ref{fig:0}(c1).
The work is supported by the EFOP-3.6.1-16-2016-00022 project. 
The project is co-financed by the European Union and the European Social Fund.
This research was supported by the National Research, Development and
Innovation Fund of Hungary, financed under the K-16 funding scheme 
Projects no.\ K 119967 and \ K 119245.
The research reported in this paper at the Budapest University of Technology and Economics has been supported by the NRDI Fund, TKP2020 IES, Grant No. TKP2020  BME-IKA-VIZ. 
The research at the University of Debrecen was supported by the Thematic Excellence Programme (TKP2020-IKA-04) of the Ministry for Innovation and Technology in Hungary.

\bibliography{/home/feri/papers/statphys_fracture}
\end{document}